\documentclass[
prc,%
10pt,%
final,%
notitlepage,%
oneside,%
twocolumn,%
nobibnotes,%
nofootinbib,% In the current version of REVTeX, when this option is on,
superscriptaddress,%
floatfix]%
{revtex4}
\usepackage{color} %{\color{red} }
\usepackage{amsfonts} 
\usepackage{amsbsy} 
\usepackage{mathrsfs}
\usepackage{graphicx}
\usepackage{comment}

\def\lsim{\mathrel{\rlap{ \lower4pt\hbox{\hskip-3pt$\sim$}}
    \raise1pt\hbox{$<$}}} %less than approx. symbol
\def\gsim{\mathrel{\rlap{ \lower4pt\hbox{\hskip-3pt$\sim$}}
    \raise1pt\hbox{$>$}}} %greater than or approx. symbol

\begin{document}
% ============================================================

\title{Kaon Modification in Pion Medium}

\author{Yu. B. Ivanov}\thanks{e-mail: yivanov@theor.jinr.ru}
\affiliation{Bogoliubov Laboratory of Theoretical Physics, JINR Dubna,
  141980 Dubna, Russia} 
\affiliation{National Research Center
  "Kurchatov Institute", 123182 Moscow, Russia}

\begin{abstract}
Kaon properties in thermal pion-nucleon medium are studied. The dense pion-nucleon medium is produced 
in high-energy heavy-ion collisions. The consideration is based on the 
chiral nucleon-kaon-pion Lagrangian. 
It is found that the pion medium 
does not produce a substantive contribution but enhances the effect of the baryon matter. 
Numerical estimate shows that this enhancement factor induced by the pion medium 
does not exceed 5\% at the freeze-out stage of heavy-ion collisions, i.e. it is small. 
However, the impact of this enhancement can be higher in actual nuclear collisions because the effect of the 
in-medium (anti)kaon modification is accumulated during the (anti)kaon evolution before the 
freeze-out when the pion density is higher. 
%
%  \pacs{25.75.-q, 25.75.Nq, 24.10.Nz} 
%	\keywords{relativistic heavy-ion collisions, hydrodynamics, directed flow}
\end{abstract}
\maketitle
% \today

% ______________________________________________________________________
%\section{Introduction} 

It is well known that kaons are modified in the baryon medium, which  
attracted attention since Kaplan and Nelson suggested possibility of kaon condensation in dense
matter \cite{Kaplan:1986yq}.
The attractive kaon-nucleon interaction may result in condensation in the interior of neutron stars, 
see review papers \cite{Watts:2016uzu,Li:2021thg,Burgio:2021vgk,Tolos:2020aln}. 
Theoretical studies of kaon production
from A+A collisions at moderately relativistic energies 
indicated that in-medium modifications of the properties of
(anti)kaons are seen in the collective flow pattern
as well as in the abundance and spectra of kaons and antikaons 
\cite{Tolos:2020aln,Hartnack:2011cn,Song:2020clw}. 

In all aforementioned studies the in-medium modifications of kaons were induced by the baryon medium. 
At higher collision energies, mesons (mostly pions) are abundantly produced. Therefore, we address the question
if the pionic medium can modify the kaon properties additionally to the baryon-induced modifications? 
This question has been already studied in Ref. \cite{Faessler:2002qb}, where the kaons were considered 
in purely pionic medium.  It was found that the main effect of that purely pionic medium is 
the kaon broadening due to rescatterings inside of the pion matter.
In particular, this broadening results in an acceleration of the $\phi \to K^+K^-$ decays. 
In the present paper the kaon modification is studied in combined pion-nucleon medium.

The present consideration follows the lines of Ref. \cite{Kolomeitsev:1995we}, 
where the influence of the Bose-Einstein pion condensate (with nonzero pion chemical potential)
on kaon properties was studied. 
In the present paper, the effect of the thermal pion-nucleon medium (with zero pion chemical potential) 
is considered. 

The starting point of the consideration is the conventional nucleon-kaon Lagrangian (see, e.g., 
Ref. \cite{Ko:1996yy}) supplemented by the chiral pion-kaon Lagrangian \cite{Belkov:1988gg}, 
which well reproduces the $\pi K$ scattering and dimeson atomic data. This Lagrangian reads 
\begin{eqnarray}
\label{L(KpiN)}
{\mathcal L}
&=&
\bar{N} (i\gamma_\mu \partial^\mu - m_N) N 
+ 
\partial_\mu \bar{K} \cdot \partial^\mu K - m_K^2 \bar{K}K
\cr
&+& 
\frac{\Sigma_{KN}}{f_K^2} [\bar{N}N\bar{K}K 
- 
\frac{3i}{8f_K^2} \bar{N}\gamma_\mu N (\bar{K} \partial^\mu K - \partial^\mu \bar{K} \cdot K)]  
%\cr
%&+&
%\partial_\mu \bar{K} \cdot \partial^\mu K 
%\left[1-\frac{\bar{K}K}{4f_\pi^2}\right]^{}_{}
%- m_K^2 \bar{K}K
%\left[1-\beta\frac{\bar{K}K}{4f_\pi^2}\right]
\cr
&+& 
%+ 
\frac{1}{2}
(\partial_\mu \boldsymbol{\pi} \cdot \partial^\mu \boldsymbol{\pi} - m_\pi^2 \boldsymbol{\pi}^2) 
%\left[1-\frac{\bar{K}K}{4f_\pi^2}\right]
%\left[1-\beta\frac{\bar{K}K}{4f_\pi^2}\right]
\cr
&+&
%\frac{1}{2f_\pi^2}
\frac{1}{4f_\pi^2}
\left[
(m_\pi^2+m_K^2)\bar{K}K\boldsymbol{\pi}^2 
- \bar{K}K\partial_\mu \boldsymbol{\pi} \cdot \partial^\mu \boldsymbol{\pi} 
\right.
\cr
&-&
\left.
 \boldsymbol{\pi}^2 \partial_\mu \bar{K} \cdot \partial^\mu K
- \bar{K}\boldsymbol{\pi} \partial_\mu K  \cdot \partial^\mu \boldsymbol{\pi} 
- (\partial_\mu \bar{K}  \cdot \partial^\mu \boldsymbol{\pi}) K  \boldsymbol{\pi} 
\right], 
\cr
&&
\end{eqnarray}
where $\boldsymbol{\pi}$ is the isotopic vector of pion fields and the kaon is the isotopic spinor
\begin{equation}
K=\left(\matrix{K^+ \cr
                K^0 \cr}\right)~~
{\rm and} ~~\bar K=(K^- ~~\bar {K^0}),
\end{equation}
$f_\pi=$ 93 MeV is the pion decay constant, $f_K$ is
the kaon decay constant that approximately equals $f_\pi$, and  $\Sigma_{KN} $ is the kaon-nucleon sigma term
\cite{Reya:1974gk}. 
The term proportional to $\Sigma_{KN}$ results from the attractive scalar interaction due
to explicit chiral symmetry breaking. 
The numerical values of the $\Sigma_{KN}$ term, derived from experimental data, vary widely \cite{Reya:1974gk}.
In practical calculations, the value of 350 MeV is usually used 
\cite{Hartnack:2011cn,Ko:1996yy}. 
The $\pi N$ interaction is omitted in this Lagrangian 
because it is irrelevant to our purposes. The chiral $\pi K$ Lagrangian contains 
additional $1/2$ factor as compared to Refs. \cite{Kolomeitsev:1995we,Belkov:1988gg}
because scalar $\boldsymbol{\pi}$ fields are used here instead of charged $\pi^+$ and $\pi^-$ in 
Refs. \cite{Kolomeitsev:1995we,Belkov:1988gg}. 

The above Lagrangian contains only $s$-wave kaon-nucleon interaction. 
Importance of $p$-wave interactions was indicated in Refs. \cite{Kolomeitsev:1995xz,Kolomeitsev:2002pg,Song:2020clw}.
Also there are more extended and sophisticated versions of the chiral $\pi K$ Lagrangian, 
see, e.g., Ref. \cite{Roessl:1999iu,Bijnens:2004bu}. 
Therefore, Lagrangian  (\ref{L(KpiN)}) can
only serve as a basis for semi-quantitative estimation of the in-medium
effects in (anti)kaon production, which is enough for our purposes.

From the Euler-Lagrange equation and using the mean-field approximation for the nucleon and pion fields 
one obtains the following Klein-Gordon equation for a kaon in the nucleon-pion medium
\begin{eqnarray}
\label{EL-eq-Kmf}
&&
\left(
1 - \frac{\langle\boldsymbol{\pi}^2\rangle}{4f_\pi^2}
\right)
(\partial_\mu \partial^\mu K + m_K^2 K) 
- \frac{\Sigma_{KN}}{f_K^2} \rho_S  K
%- \frac{\Sigma_{KN}}{f_K^2} \langle\bar{N}N\rangle K
\cr
&+&
\frac{3i}{4f_K^2} J_\mu  \partial^\mu K   
%\cr
%&-&
- \frac{1}{4f_\pi^2}
\langle\partial_\mu \boldsymbol{\pi}^2\rangle \cdot \partial^\mu K
\cr
&-&
%\frac{1}{2f_\pi^2}
\frac{1}{4f_\pi^2}
%\underbrace{
\left[
m_\pi^2 \langle\boldsymbol{\pi}^2\rangle 
+ \langle\boldsymbol{\pi} \partial_\mu\partial^\mu \boldsymbol{\pi} \rangle    
\right]
%}_{=0 \;\;\mbox{\scriptsize if pions are ideal gas}}
 K
= 0, 
\end{eqnarray}
where 
\begin{eqnarray}
\label{JN}
J_\mu =\langle\bar{N}\gamma_\mu N\rangle
\\
\label{rhoS}
\rho_S = \langle\bar{N}N\rangle
\end{eqnarray}
are the nucleon current and scalar density, respectively. 
In the homogeneous and stationary medium 
$$
\partial_\mu \langle\boldsymbol{\pi}^2\rangle = 0. 
$$

The term $[...]$ in Eq. (\ref{EL-eq-Kmf}) can be estimated by means of the Euler-Lagrange equation for pions. 
Then it is 
$$
%\frac{1}{2f_\pi^2}
\frac{1}{4f_\pi^2}
\left[
m_\pi^2 \langle\boldsymbol{\pi}^2\rangle 
+ \langle\boldsymbol{\pi} \partial_\mu\partial^\mu \boldsymbol{\pi} \rangle    
\right]
\propto 
\frac{\langle\bar{K}K\rangle}{4f_\pi^2}  \frac{\langle\boldsymbol{\pi}^2\rangle}{4f_\pi^2}  ,  
$$
i.e. contains additional smallness
$$
%\frac{\langle\boldsymbol{\pi}^2\rangle}{4f_\pi^2}   < 1,  
%\hspace*{9mm}\mbox{and}\hspace*{9mm}
\frac{\langle\bar{K}K\rangle}{\langle\boldsymbol{\pi}^2\rangle} \ll 1.  
$$
We also could add the pion nonlinear (Weinberg) term into the Lagrangian (\ref{L(KpiN)}) \cite{Kolomeitsev:1995we}
$$
-\frac{1}{8f_\pi^2}
(2\boldsymbol{\pi}^2 \partial_\mu \boldsymbol{\pi} \cdot \partial^\mu \boldsymbol{\pi} - m_\pi^2 \boldsymbol{\pi}^4
) 
$$
similarly to Ref. \cite{Kolomeitsev:1995we}. 
Then it would be 
$$
%\frac{1}{2f_\pi^2}
\frac{1}{4f_\pi^2}
\left[
m_\pi^2 \langle\boldsymbol{\pi}^2\rangle 
+ \langle\boldsymbol{\pi} \partial_\mu\partial^\mu \boldsymbol{\pi} \rangle    
\right]
\propto 
\left(\frac{\langle\boldsymbol{\pi}^2\rangle}{4f_\pi^2} \right)^2,   
$$
i.e. it also contains extra smallness. 
Thus,  the term $[...]$ in Eq. (\ref{EL-eq-Kmf}) can safely neglected.

Finally, in the rest frame of homogeneous and stationary medium, i.e. $\boldsymbol{J}=0$,  we arrive at 
\begin{eqnarray}
\label{EL-eq-Kmff}
%&\hspace*{-7mm}&
%&& 
\partial_\mu \partial^\mu K + m_K^2 K 
%\cr
%&\hspace*{-7mm}+& 
%&+& 
+ F_\pi 
\left[
%+ 
\frac{3i}{4f_K^2} \rho_N  \partial_t K   
- \frac{\Sigma_{KN}}{f_K^2} \rho_S  K
\right]
= 0, 
\end{eqnarray}
where $\rho_N$ is the nucleon density in the rest frame, $J_\mu =(\rho_N,0,0,0)$, and 
\begin{eqnarray}
\label{Fpi}
F_\pi = \left(
1 - \frac{\langle\boldsymbol{\pi}^2\rangle}{4f_\pi^2}
\right)^{-1}. 
\end{eqnarray}
Thus, we arrive at the well-known equation for the kaon in the nucleon medium \cite{Ko:1996yy} but with 
additional enhancement factor $F_\pi$ that results from  
the pion medium. Similar manipulations for anti-kaons result to the same equation except for 
the opposite sign in front of the $\rho_N  \partial_t \bar{K}$ term.

The (anti)kaon dispersion relation in nuclear-pion matter is then given by
\begin{eqnarray}
\omega^2
%({\bf k},\rho _B) 
=m_K^2+{\bf k}^2 -\frac{F_\pi\Sigma_{KN}}{f_K^2} \rho_S %+ 
\pm
\frac{3F_\pi}{4f_K^2}
\omega \rho_N,
\label{DIS}
\end{eqnarray}
where the upper (lower) sign in front of the $\omega \rho_N$ term
corresponds to the (anti)kaon and $(\omega,{\bf k})$ is the four-momentum of the (anti)kaon,
The quantity $\langle\boldsymbol{\pi}^2\rangle$ reads
\begin{eqnarray}
\label{Sdis_pi}
\langle\boldsymbol{\pi}^2\rangle = \frac{\rho_S^\pi}{2m_\pi} = 
g_\pi \int \frac{d^3 k}{(2\pi)^3 2\omega_\pi(k)} 
\frac{1}{\exp (\omega_\pi/T) -1}
\end{eqnarray}
where $\rho_S^\pi$ is the scalar pion density, $g_\pi=3$ is the pion degeneracy factor, and  $T$ is the temperature of the medium.

From the dispersion relation, the (anti)kaon energy in medium can be 
obtained
\begin{eqnarray}
\label{omek}
&\hspace*{-7mm}&
\omega({\bf k})=
\sqrt{m_K^2+{\bf k}^2-\frac{F_\pi\Sigma_{KN}}{f_K^2}\rho_S
+\left(\frac{3F_\pi\rho_N}{8f_K^2}\right)^2}
%
%\left[m_K^2+{\bf k}^2-\frac{F_\pi\Sigma_{KN}}{f_K^2}\rho_S
%+\left(\frac{3F_\pi\rho_N}{8f_K^2}\right)^2\right]^{1/2}
%\!\!\!\! \pm
\! \pm
\frac{3F_\pi\rho_N}{8f_K^2},
\cr
&\hspace*{-17mm}&
\label{DIS1}
\end{eqnarray}
where again the upper(lower) sign refers to $K$($\bar{K}$). 
The nucleon scalar and number densities, $\rho_S$ and $\rho_N$, respectively, 
are usually replaced by respective baryon densities in actual calculations. 
As seen, the pion medium 
does not produce a substantive contribution but enhances the effect of the baryon matter.

To quantitatively estimate the enhancement factor $F_\pi$ due to the pion medium, 
use is made of the freeze-out temperatures and baryon chemical potentials ($\mu_B$) deduced
from experimental data with the statistical model \cite{Cleymans:2005xv,Karsch:2010ck}
$$
T(\mu_B) = a-b\mu_B^2-c\mu_B^4
$$
where $a=0.166$ GeV, $b=0.139$ GeV$^{-1}$, $c=0.053$ GeV$^{-3}$, and
%where  = 0.166  GeV, $b$ = 0.139 GeV$^{−1}$, $c$ = 0.053 GeV$^{−3}$, and %$a$
$$
\mu_B(\sqrt{s_{NN}}) =
\frac{d}{1 + e\sqrt{s_{NN}}}
$$
with $d=1.308$ GeV and $e=0.273$ GeV$^{-1}$. 
%with $d$ = 1.308 GeV and $e$ = 0.273 GeV$^{−1}$. 
Here, $\sqrt{s_{NN}}$ is the center-of-mass energy of the nuclear collision (in GeV).

\begin{figure}[!h]
%\vspace*{-14mm}
\includegraphics[width=6.6cm]{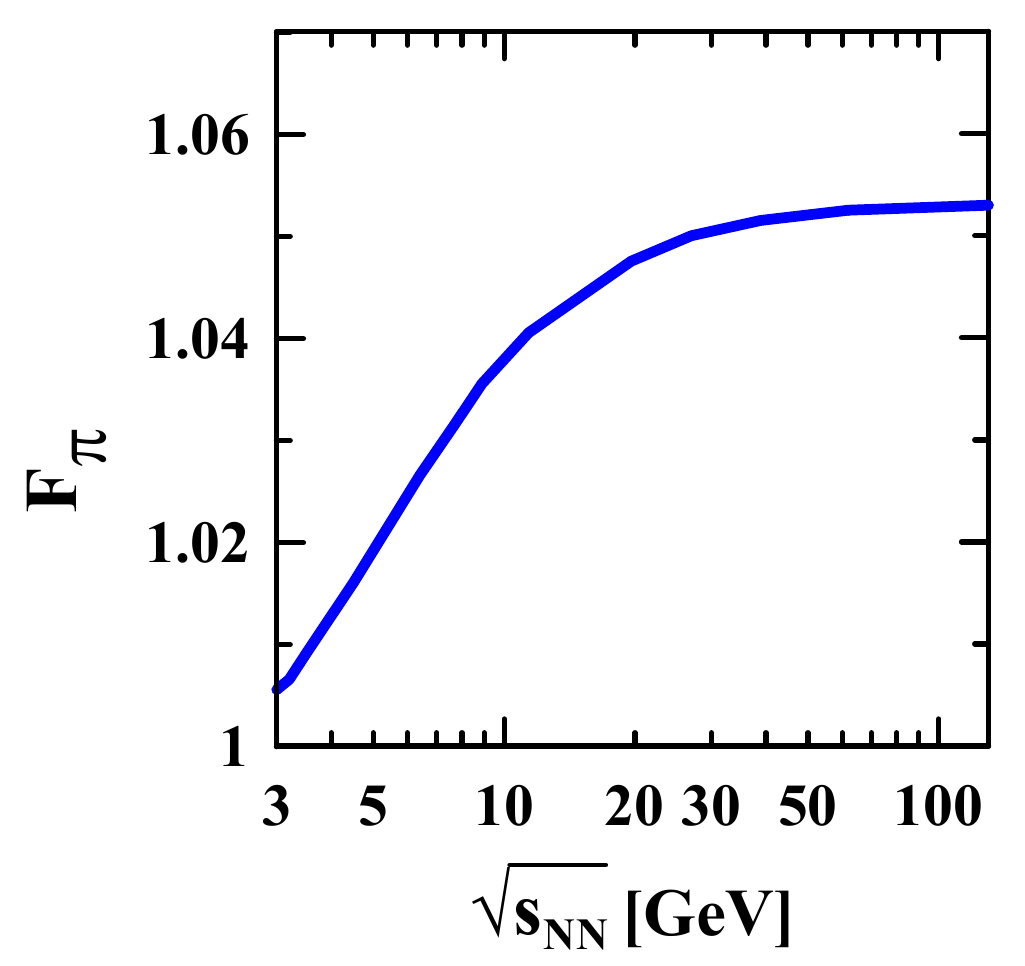}
 \caption{%(Color online)
Enhancement factor $F_\pi$ induced by the pion medium at the freeze-out stage as a function 
the center-of-mass energy of the nuclear collision $\sqrt{s_{NN}}$.  
}
\label{fig:K-pi-interaction}
\end{figure}

The result of numerical calculation of the enhancement factor $F_\pi$ induced by the pion medium 
at the freeze-out stage as a function 
the center-of-mass energy of the nuclear collision is presented in Fig. \ref{fig:K-pi-interaction}. 
As seen, it does not exceed 5\%, i.e. it is small. This retroactively confirms the approximations made
at derivation of Eq. (\ref{omek}). 
However, the impact of this enhancement can be higher in actual nuclear collisions because the effect of the 
in-medium (anti)kaon modification is accumulated during the (anti)kaon evolution before the 
freeze-out when the pion density is higher.

Enlightening  discussions with  D.N. Voskresensky and E.E. Kolomeitsev are gratefully acknowledged.
This work was carried out using computing resources of the federal collective usage center ``Complex for simulation and data processing for mega-science facilities'' at NRC "Kurchatov Institute" \cite{ckp.nrcki.ru}.

% ________________________________________________________________

%%%%%%%%%%%%%%%%%%%%%%%%%%%%%%%%%%%%%%%%%%%%%%%%%%%%%
\end{document}